\def\ro{{\it ROSAT}}
\def\asca{{\it ASCA}}
\def\cgro{{\it CGRO}}

\def\lum{ergs~s$^{-1}$}
\def\hst{{\it HST}}

\def\ep{$e^{\pm}$}
\def\gr{$\gamma$-ray}

\documentstyle[12pt,aaspp4]{article}
\begin{document}  
\title{Models for X-Ray Emission from Isolated Pulsars}
\author{F.~Y.-H.~Wang, M.~Ruderman, J.~P.~Halpern, and T.~Zhu}
\affil{Columbia Astrophysics Laboratory, Columbia University,\\
550 West 120th Street, New York, NY 10027 \\
fyw@orphee.phys.columbia.edu}

\begin{abstract}

A model is proposed for the observed combination of power-law and thermal
emission of keV X-rays from rotationally powered pulsars.  For \gr\
pulsars with accelerators very many stellar radii above the neutron star
surface, 100 MeV curvature \gr s from $e^{-}$ or $e^{+}$ flowing starward
out of such accelerators are converted to \ep\ pairs on closed field lines
all around the star.  These pairs strongly affect X-ray emission from near
the star in two ways.  (1) The pairs are a source of synchrotron emission
immediately following their creation in regions where $B \sim 10^{10}$ G. 
This emission, in the photon energy range 0.1 keV $\lesssim E_{\rm X}
\lesssim$ 5 MeV, has a power-law spectrum with energy index 0.5 and X-ray
luminosity that depends on the backflow current, and is typically $\sim
10^{33}$ \lum.  (2) The pairs ultimately form a cyclotron resonance
``blanket'' surrounding the star except for two holes along the open field
line bundles which pass through it.  In such a blanket the gravitational
pull on \ep\ pairs toward the star is balanced by the hugely amplified
push of outflowing surface emitted X-rays wherever cyclotron resonance
occurs. Because of it the neutron star is surrounded by a leaky
``hohlraum'' of hot blackbody radiation with two small holes, which
prevents direct X-ray observation of a heated polar cap of a \gr\ pulsar. 
Weakly spin-modulated radiation from the blanket together with more
strongly spin-modulated radiation from the holes through it would then
dominate observed low energy (0.1--10 keV)  emission.  For non-\gr\
pulsars, in which no such accelerators with their accompanying extreme
relativistic backflow toward the star are expected, optically thick \ep\
resonance blankets should not form (except in special cases very close to
the open field line bundle).  From such pulsars blackbody radiation from
both the warm stellar surface and the heated polar caps should be directly
observable.  In these pulsars, details of the surface magnetic field
evolution, especially of polar cap areas, become relevant to observations. 
The models are compared to X-ray data from Geminga, PSR 1055--52, PSR
0656+14, PSR 1929+10, and PSR 0950+08.

\end{abstract}


\section{Introduction}

X-ray emission at keV energies from pulsars may come from several sources: 
(a) thermal cooling from the entire stellar surface of deep internal
heating source (e.g., residual heat from the violent stellar birth,
glitches, and magnetic flux tube motion through the electron-proton sea of
the core); (b) polar cap heating from an inflow of extremely relativistic
$e^{-}$ or $e^{+}$ that have passed through an accelerator on the open
field line bundle;  (c)  synchrotron radiation from \ep\ pairs, created in
the near magnetosphere above the stellar surface at altitude $\sim 3 R$;
(d) inverse Compton scattering of radio photons by extremely relativistic
$e^{-}$ or $e^{+}$ inflow or outflow on the open field lines of the
pulsar.  We shall consider here interpretation of X-ray observations from
solitary (isolated) pulsars in terms of these assumed sources. 

In \S 2 we summarize X-ray emission and pulsar properties: present data
seem adequate for differentiating among possible sources.  A dichotomy
exists between the data from \gr\ pulsars and ``ordinary'' ones (those
that are not observed to have strong \gr\ emission).  The former show a
power-law spectrum for their hard X-rays (1--10 keV) and no evidence for
observation of the expected heated polar cap emission.  The latter do not
show the strong strong power-law emission, but heated polar cap emission
does appear to have been observed.  We interpret these differences as a
consequence of \ep\ production in the near magnetospheres of \gr\ pulsars
but not in those of ordinary ones.

The locations of major particle accelerators in pulsar magnetospheres and
the flow of particles through them may be different from those which power
the much lower energy radio emission. We discuss relevant features of
magnetospheric accelerators in \S 3. 

Power-law emission in the keV range with the observed spectral extent,
intensity, and angular distribution seem to be a natural consequence of
the outer magnetospheric accelerator models, but not of polar-cap ones. 
Modelling the source of power-law emission is the subject of \S 4.

Copious \ep\ production on closed field lines above the neutron star
surface, no matter what its origin, would sustain a cyclotron-resonant
X-ray reflecting mirror there (referred to as a ``blanket'').  This
blanket would reflect X-rays from a hot polar cap in all directions except
in a narrow cone through either of the two holes in the blanket where the
open field line bundle penetrates through it.  These cyclotron-resonant
blankets are considered in \S 5.  Such thick blankets should only exist on
the closed field lines above the surfaces of \gr\ pulsars of which the
hard X-ray power-law spectrum is from the synchrotron radiation of \ep\
pairs made in the near-magnetosphere.  Direct observation of X-rays from
heated polar caps through the holes of the open field line bundle would be
expected to be rare: most of these X-rays from a hot polar cap would be
reflected back toward the star by the blanket. 

In \S 6 we discuss features of individual pulsars that need special
consideration.  The \gr\ pulsar Geminga has a soft X-ray (0.1--1 keV) 
luminosity that is not as large as that from the heating of a polar cap
expected from the down-flowing extremely relativistic $e^{-}$ or $e^{+}$
that strike it.  A possible mechanism for reducing the power of that
inflow by inverse Compton scattering of inflowing extremely relativistic
$e^{-}$ ($e^{+}$)  by radio photons emitted near the polar cap is
presented.

The presence or absence of a strong blackbody component of hard X-rays
depends on the formation of the blanket on closed field lines of the near
magnetosphere. For most non-\gr\ pulsars, such blankets would not be
expected, and direct observation of their hot polar caps is possible.  In
\S 7 we consider the effective areas of the X-ray emission and
luminosities of hot polar caps. 

If the blanket is present, soft X-rays (0.1--1 keV) can escape from the
star either by penetrating through the blanket itself, which gives a
weakly modulated light curve, or by leaking out through the holes of the
blanket on open field lines, which gives a strongly modulated one.  The
combination of these two sources might explain the pulsed fraction
observed in various pulsars.  Because of uncertainties with luminosity and
density distribution, a quantitative calculation is not yet available.  We
only outline this concept in \S 8. 

\section{X-ray Observations}

Following the detection by the {\it Einstein} satellite of several
rotationally powered pulsars at X-ray wavelengths, their surface emission
and magnetospheric processes have been studied using spectra and light
curves obtained by the {\it R\"{o}ntgen Satellite} (\ro) and the {\it
Advanced Satellite for Cosmology and Astrophysics} (\asca).  About two
dozen isolated pulsars have been detected by these instruments (Finley,
\"{O}gelman, \& K{\i}z{\i}lo\u{g}lu 1992; \cite{hal92}; \cite{oge93}; 
Yancopoulos, Hamilton, \& Helfand 1994; \cite{man94}).  X-ray spectra are
usually fitted to either a thermal blackbody shape or a nonthermal
power-law one. 

Geminga is a radio-quiet pulsar with period $P=0.237$ s, surface magnetic
field $B_{\rm p} \sim 1.6 \times 10^{12}$ G, and characteristic age $\tau
= P/2\dot{P} = 3.4 \times 10^{5}$ years (\cite{hal92}; \cite{ber92}). 
Halpern and Ruderman (1993) fitted the \ro\ PSPC spectrum of Geminga with
a two-component model, either two blackbodies, or a blackbody plus a
power-law.  Both fits gave a satisfactory description of the spectrum. 
This ambiguity was resolved by combining \ro\ and \asca\ data: the bulk of
the soft X-ray flux is a blackbody, and the hard X-ray spectrum is a power
law (\cite{hal97}).  The soft X-ray component is parameterized as a
blackbody with $T = 5.77^{+0.37}_{-0.46} \times 10^{5}$ K and a bolometric
luminosity $1.47 \times 10^{31}$ \lum; the hard X-ray spectrum is fitted
by a power law with energy index $0.47^{+0.25}_{-0.23}$ and luminosity
$8.13 \times 10^{29}$ \lum (Figure 1).  The luminosities were calculated
from the parallax distance of 160 pc from the \hst\ measurements
(\cite{car96}).

\begin{figure}
\plotone{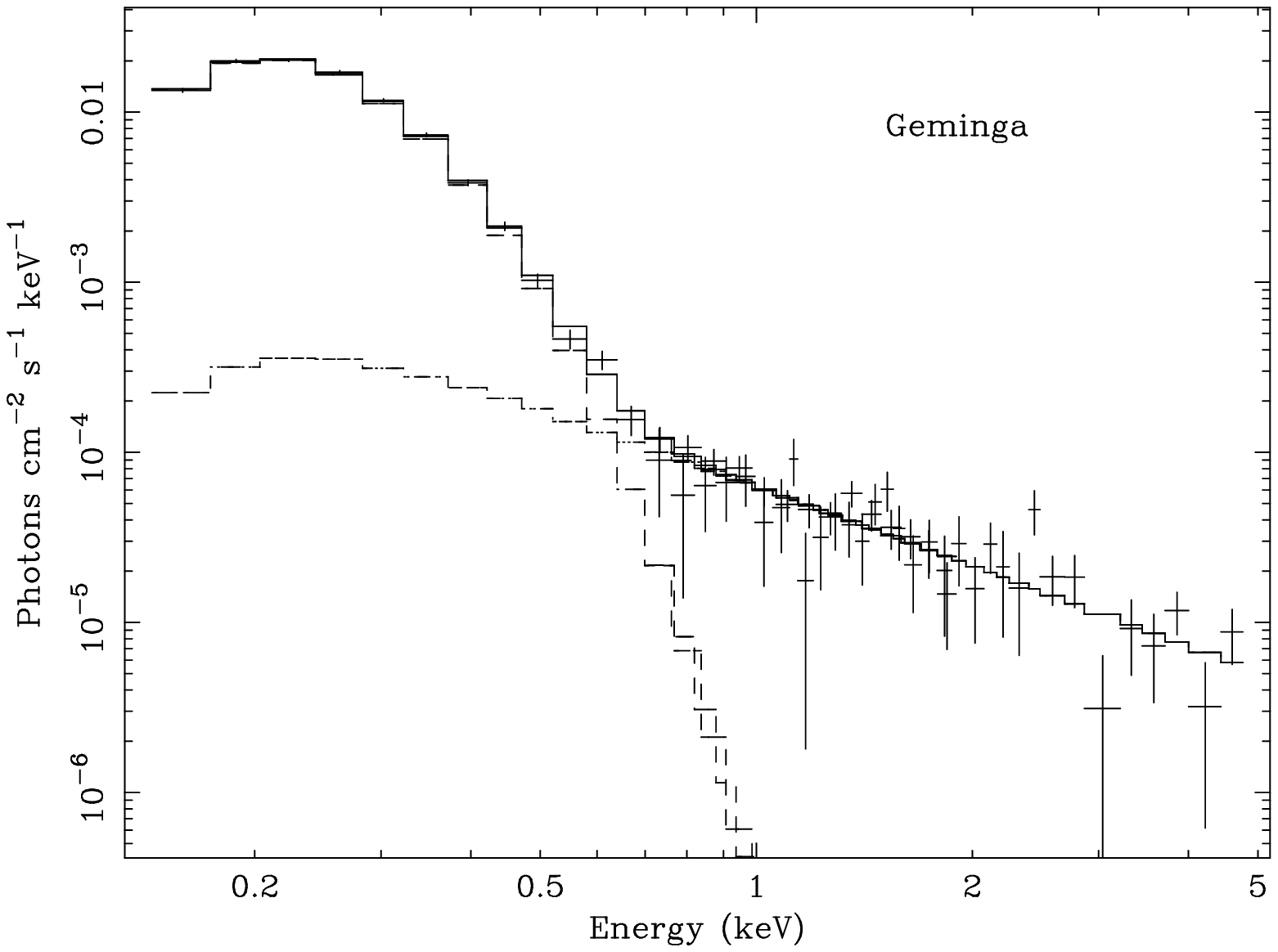}
\caption{Fit of the blackbody plus power-law model to the
\asca\ and \ro\ spectra of Geminga. \label{fig1}}
\end{figure}

\placefigure{fig1}

PSR 1055--52 and PSR 0656+14 resemble Geminga in many respects: periods,
loss rate of rotational energy, characteristic age, etc. (Table 1).  They
have also been fitted with a two-component model (\cite{oge93}; 
\cite{gre96}; Possenti, Mereghetti, \& Colpi 1996). The soft X-ray
component has generally been interpreted as thermal emission from all or a
part of the surface of the neutron stars.  However, the source of the
harder X-ray component has not been determined.  \"{O}gelman and Finley
(1993) analyzed the \ro\ data from PSR 1055--52;  their results were
fitted to a blackbody together with a power law of energy index 0.4--0.5.
Although Greiveldinger et al. (1996)  favored a blackbody model to
describe the hard X-ray component for a joint fit of \ro\ and \asca\ data,
our own reexamination of the same \asca\ data shows that a power law with
energy index $0.5 \pm 0.3$ and luminosity $1.5 \times 10^{30}$ \lum
(calculated at $d=500$ pc) fits the spectrum equally well (Figure 2).  PSR
1055--52 possesses a large pulsed fraction, 0.73 $\pm$ 0.33 at higher
X-ray energies (\cite{oge93}). Because of the gravitational bending
effect, pulsed fraction due to heated polar cap can rarely exceed $50 \%$,
and a high pulsed fraction seems more easily compatible with a strongly
beamed nonthermal hard X-ray component.  [See simulations by Possenti et
al.~(1996)  and Wang \& Halpern (1997).] The soft X-ray component has
temperature $7.9^{+0.6}_{-1.0} \times 10^{5}$ K and bolometric luminosity
$2.2 \times 10^{32}$ \lum.  Greiveldinger et al.  (1996) also reported a
joint analysis of \ro\ and \asca\ data for PSR 0656+14.  They required two
blackbodies and a power law to fit the spectrum.  Our own analysis
verified the double blackbody model but failed to confirm the extra
power-law component.  The soft X-ray component has temperature
$8.1^{+0.5}_{-1.4} \times 10^{5}$ K and bolometric luminosity $4.3 \times
10^{32}$ \lum; the hard X-ray component has temperature $1.7^{+0.2}_{-0.2}
\times 10^{6}$ K and a bolometric luminosity $2.6 \times 10^{31}$ \lum. 
The luminosities are calculated at an estimated distance of 400 pc. 

\begin{figure}
\plotone{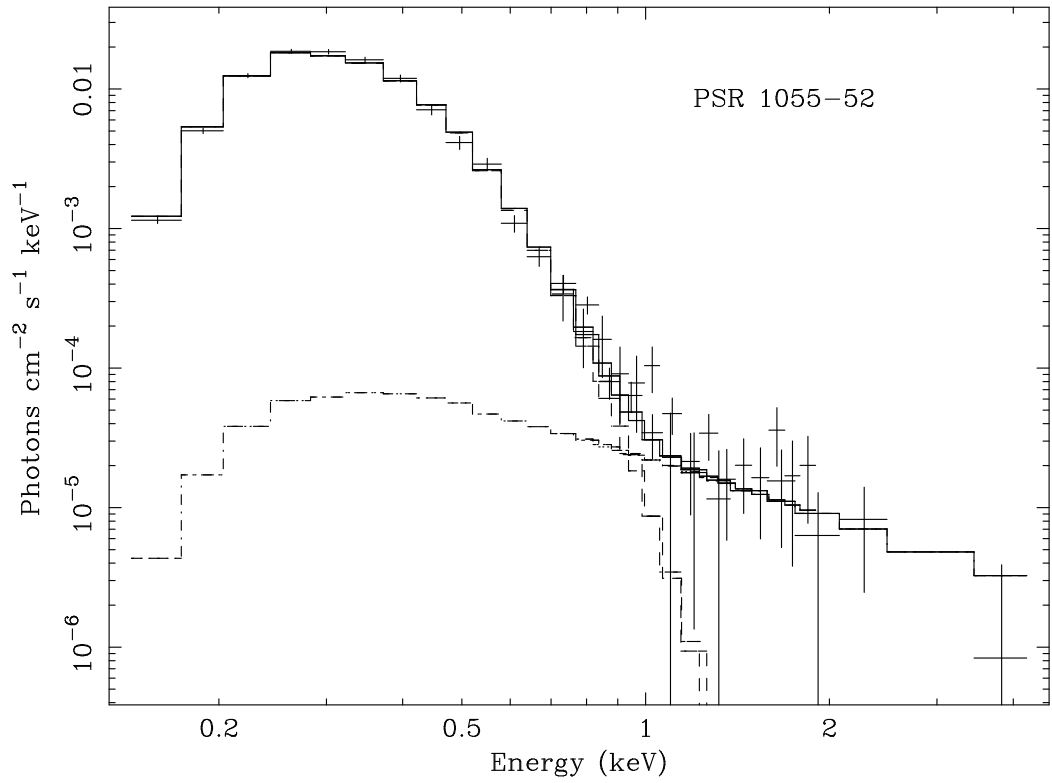}
\caption{Fit of the blackbody plus power-law model to the 
\asca\ and \ro\ spectra of PSR 1055--52. \label{fig2}}
\end{figure}

\placefigure{fig2}

The older pulsars PSR 1929+10 and PSR 0950+08 have similar period, and
roughly comparable characteristic ages, spin-down powers, and surface
magnetic fields (Table 1).  According to the cooling curves (\cite{pag92}; 
\cite{lat91}), the X-ray luminosity of a pulsar whose age exceeds $10^{6}$
years decreases sharply, and the typical energy of thermal photons from
such an object is too low to create pairs in collisions with GeV photons. 
X-rays are detected from both pulsars by \ro\ (\cite{yan94}; 
\cite{man94})  and \asca\ (\cite{wan97}).  The spectra from both pulsars
are fitted to a single blackbody spectrum from a small area.  PSR 1929+10
is best fitted with temperature $(5.14 \pm 0.53)  \times 10^{6}$ K and a
luminosity $1.54 \times 10^{30}$ \lum\ after adopting the distance 250 pc
of Yancopoulos et al. (1994). PSR 0950+09 has temperature $(5.70 \pm 0.63) 
\times 10^{6}$ K and a luminosity $4.67 \times 10^{29}$ \lum, at the
parallax distance of 125 pc (\cite{gwi86}). 

At present, about twenty pulsars have been detected in soft X-rays, but
few of them have sufficient photons for comprehensive spectral and
temporal analysis.  The above five objects, observed in the X-ray energy
range 0.1--10.0 keV, represent pulsars of various classes.  They all have
periods of about 0.3 s. Geminga, PSR 1055--52, and PSR 0656+14 have
characteristic ages $\tau \sim 10^{5}$ years, whereas PSR 1929+10 and PSR
0950+08 have $\tau \sim 10^{7}$ years. Strong pulsed \gr\ emission is
detected from Geminga (\cite{ber92})  and PSR 1055--52 (\cite{fie93}), but
such radiation is still uncertain for PSR 0656+14 (\cite{ram96}).  While
Geminga, PSR 1055--55, and PSR 0656+14 have typical dipole component
magnetic fields $B_{\rm p} \gtrsim 10^{12}$ G, PSR 1929+10 and PSR 0950+08
have weaker fields ($B_{\rm p} < 10^{12}$ G).  Properties of these five
pulsars are summarized in Table~1.  Luminosities have a great uncertainty
associated with the measurements of distance.  We observe thermal soft
X-ray emission from Geminga, PSR 1055--52, and PSR 0656+14, and the
emitting area is comparable to the entire surface of a neutron star.  A
thermal hard X-ray component is observed from PSR 1929+10, PSR 0950+08,
and PSR 0656+14 from an area which occupies only a small fraction of the
neutron star surface. A power-law spectrum in hard X-rays is observed from
the \gr\ pulsars Geminga and PSR 1055--52.  Light curves of the hard
X-rays of both \gr\ pulsars have pulsed fractions exceeding 50\%, which we
interpret as an indicator of nonthermal emission. 

\placetable{tbl-1}

\section{Particle Flows in Pulsar Magnetospheres}

The Energetic Gamma Ray Experiment Telescope (EGRET) on the {\it Compton
Gamma Ray Observatory} (\cgro) has detected six pulsars with strong \gr\
emission above 100 MeV: Crab, Vela, Geminga, PSR 1706--44, PSR 1055--52,
and PSR 1951+32.  Similarities among the \gr\ spectra and most of the
light curves suggest that a similar powerful accelerator is operating in
each of their magnetospheres.

Models for magnetospheric accelerator gaps fall into two main classes. 
For a polar-cap accelerator (\cite{rud75}, RS hereafter; \cite{dau82}),
generally assumed to power the radio emission of a pulsar, the
acceleration of primary particles takes place relatively near the surface
of a neutron star.  In the central dipole approximation for the stellar
magnetic field, all the radiation and pair production resulting from
primaries which have passed through such accelerators would be restricted
to be within the open field-line bundle that links the polar cap to the
light cylinder ($r_{\rm LC} \equiv c/\Omega$, in which $\Omega=2 \pi
P^{-1}$). Outer-magnetospheric accelerators have been proposed for the
emission of the energetic photons from \gr\ pulsars (Cheng, Ho, \&
Ruderman 1986a,b, hereafter CHRa, CHRb).  These accelerators could achieve
much higher potential drops along {\bf B},
\begin{equation} 
\Delta V 
\sim \frac{\Omega^{2} B_{\rm p} R^{3}}{c^{2}} \sim 10^{14} \ \rm V,
\end{equation} 
with $B_{\rm p}$ the surface dipole field, than can polar cap
accelerators.  Pair production within such accelerators (e.g. from
accelerator produced GeV \gr s colliding with keV X-rays from the stellar
surface or its neighborhood) will result in a net flow of primary
$e^{-}$($e^{+}$) out of the star being balanced by an inward flow of
$e^{+}$($e^{-}$) from the starward end of the accelerator.  Because of
extensive \ep\ flows generated by both polar cap and outer-magnetospheric
accelerators, it is difficult to see how both could survive on the same
field lines: pairs from one would be expected to quench the other. 
However, a polar cap might be linked to both kinds of accelerator as long
as the field lines through each are different [e.g. upward curving with a
net positive (negative) charge near the light cylinder for one and
downward curving negative (positive) for the other, or lines passing
through the charge reversing surface $\hat{\Omega} \cdot {\bf B} =0$ for
the one but not for the other, etc.]. 

Arguments for important outer-magnetospheric accelerators as well as polar
cap ones in \gr\ pulsars but not in ordinary ones include the following: 
(a)  apparently different death lines for the two families (\cite{ch93a}); 
for example ordinary pulsar radio emission appears to cease when polar cap
accelerator \ep\ production is no longer possible while \gr\ pulsar
emission seems to be quenched when putative outer-magnetospheric
accelerators would no longer be able to sustain the mechanism of collision
between GeV \gr s and keV X-rays; (Thus, Geminga and PSR 1055--52 could
maintain outer-magnetospheric accelerators, but PSR 1929+10 and PSR
0950+08 could not;  the former are \gr\ pulsars, while the latter are
not.)  (b) strong optical emission coincident with X-ray emission from the
Crab pulsar which is difficult to explain unless its source is synchrotron
radiation in the outer magnetosphere; (c)  escape of 10 GeV \gr s from PSR
1951+32 and PSR 1706--44 which seems more difficult to accomplish if these
\gr s originate in the neighborhood of strong polar cap fields; (d) the
\gr\ luminosity of these \gr\ pulsars with $P \gtrsim 10^{-1} \ \rm
s^{-1}$ ($L_{\gamma} \sim 10^{34}$ \lum).  From an outer-magnetospheric
accelerator, because almost the full potential drop $\Delta V \approx
10^{14}$ V will be radiated away by each electron in the accelerator, we
indeed expect just such a luminosity,
\begin{equation}
L_{\gamma} \sim  \dot{N}_{0} e \Delta V \sim 10^{34} \
{\rm ergs \ s^{-1}},
\end{equation}
for $\dot{N}_{0}$ the maximum current through the accelerator (i.e.
the Goldreich-Julian current)
\begin{equation}
\dot{N}_{0} \sim \frac{\Omega^{2}B_{\rm p}R^{3}}{ec} \sim 10^{32} \ \rm
s^{-1}. \label{eq:GJ}
\end{equation}

We assume that outer-magnetospheric accelerators---or at least their
starward flow of extreme relativistic leptons---are present in \gr\
pulsars. [We note, however, that the outer-magnetospheric accelerator
model of Romani (1996) and co-workers does not include any such symmetry
between outward $e^{-}$($e^{+}$) and inward $e^{+}$($e^{-}$) flow.] In the
next section, an interpretation of X-ray data from \gr\ pulsars provides
more evidence favoring outer-magnetospheric accelerators as a source of
extremely relativistic starward particle flow.

\section{X-rays from \gr\ Pulsars}

The \gr\ pulsars Geminga, PSR 1055--52, and PSR 1951+32 all have a
power-law hard X-ray component with energy index $\sim$ 0.5.  If
synchrotron radiation in the magnetosphere is responsible for this
emission, an outer-magnetospheric model can explain its slope and
intensity.  The polar-cap accelerator model, on the other hand, has
difficulties accounting for this component, as we describe below.

\subsection{Spectra} 

The energy loss rate $\dot{E}$ of extremely relativistic charged particles
flowing along curved paths is
\begin{equation}
\dot{E}=\dot{\gamma} m c^{2} = -\frac{2}{3} \frac{e^{2}c}{s^{2}} 
\gamma^{4},
\end{equation}
with $s$ the radius of curvature.  For \gr\ pulsars powered by
outer-magnetospheric accelerators, the number of particles flowing through
an accelerator along the open field line bundle is limited by the
Goldreich-Julian current.  Although an outer-magnetospheric accelerator
might sustain a potential difference along open field lines as high as
$\Delta V \sim 10^{14}$ V, radiation reaction limits the energy of primary
particles $E_{\rm p}$ (CHRb) 
\begin{equation}
\gamma_{1} \equiv \frac{E_{\rm p}}{m c^{2}} \leq \left(\frac{{\bf E
\cdot \hat{B}}}{e} s^{2}\right)^{1/4}
\sim 3 \times 10^{7},
\end{equation}
For a dipole field, $s \cong (rc/\Omega)^{1/2}$, with $r$ the radial
distance. When primary particles flowing starward from an
outer-magnetospheric accelerator approach the polar cap, $\gamma$
satisfies
\begin{equation}
\frac{1}{\gamma^{3}} - \frac{1}{\gamma_{1}^{3}} \sim \frac{2 \Omega e^{2}}
{m c^{3}} \ln \left(\frac{r_{\rm LC}}{r}\right), \label{eq:res}
\end{equation}
with $r \cong ct$.  If $r \ll r_{\rm LC}$, $\gamma$ is
insensitive to $r$ and $\gamma_{1}$; it remains approximately constant at
\begin{equation}
\gamma \sim \left(\frac{2 \Omega e^{2}}{mc^{3}}\ln \frac{r_{\rm LC}}{r}
\right)^{-1/3}
\sim 10^{7}. \label{eq:gamm}
\end{equation}
Each such $e^{-}$($e^{+}$) produces curvature radiation with  
\begin{equation}
E_{\gamma} \sim \hbar \gamma^{3} \left(\frac{c}{s}\right) \sim 100 \ \rm MeV. 
\end{equation}
The number of such $\gamma$-rays radiated during the traversal along $\bf
B$ a distance equal to the local radius of curvature is
\begin{equation}
N_{\gamma} \sim \frac{\dot{E} s}{E_{\gamma} c} \sim \frac{\gamma
e^{2}}{\hbar c} \sim 10^{5}. \label{eq:multi}
\end{equation}

Photons from curvature radiation are initially emitted almost tangent to
the local $\bf B$ within an angle $\gamma^{-1}$; after they proceed a
distance $l$, they make an angle $\phi$ with the magnetic field line,
\begin{equation}
\phi \sim l/s.
\end{equation}
These $10^{2}$ MeV $\gamma$-ray photons are converted into \ep\ pairs
at a significant rate as soon as they cross magnetic field lines that
satisfy (RS)
\begin{equation}
\frac{E_{\gamma}\sin\phi}{2mc^{2}}\frac{B_{12}}{2}
\geq 1, \label{eq:paircvt}
\end{equation}
The angle $\phi$ becomes the pitch angle of the \ep\ pair trajectories. 
With an outer-magnetospheric accelerator model for starward primary
particle flow, most 100 MeV photons will be converted on closed field
lines at $B \sim 10^{10}$ G, and $\phi \sim 90^{\circ}$.  Each electron
and positron carries about half the parent $\gamma$-ray energy, and
quickly loses its momentum perpendicular to the local magnetic field by
synchrotron radiation at the rate
\begin{equation}
\frac{1}{\tau_{\rm s}} \approx
\left(\frac{\gamma_{\perp}}{\gamma_{\parallel}}\right)\,\left(\frac{e^{2}}
{mc^{3}}\right)\,\omega_{B}^{2} \sim 10^{11} \ \rm s^{-1},
\end{equation}
with
\begin{equation}
\gamma_{\perp} \approx \gamma \sin \phi,
\end{equation}
\begin{equation}
\gamma_{\parallel}=\frac{\gamma}{\gamma_{\perp}} \approx \frac{1}
{\sin \phi},
\end{equation}
\begin{equation}
\omega_{B} \equiv \frac{eB}{mc}.
\end{equation}
The distance that a charged particle can propagate before losing most of
its energy is $\sim 0.3$ cm (Figure 3). 

\begin{figure}
\plotone{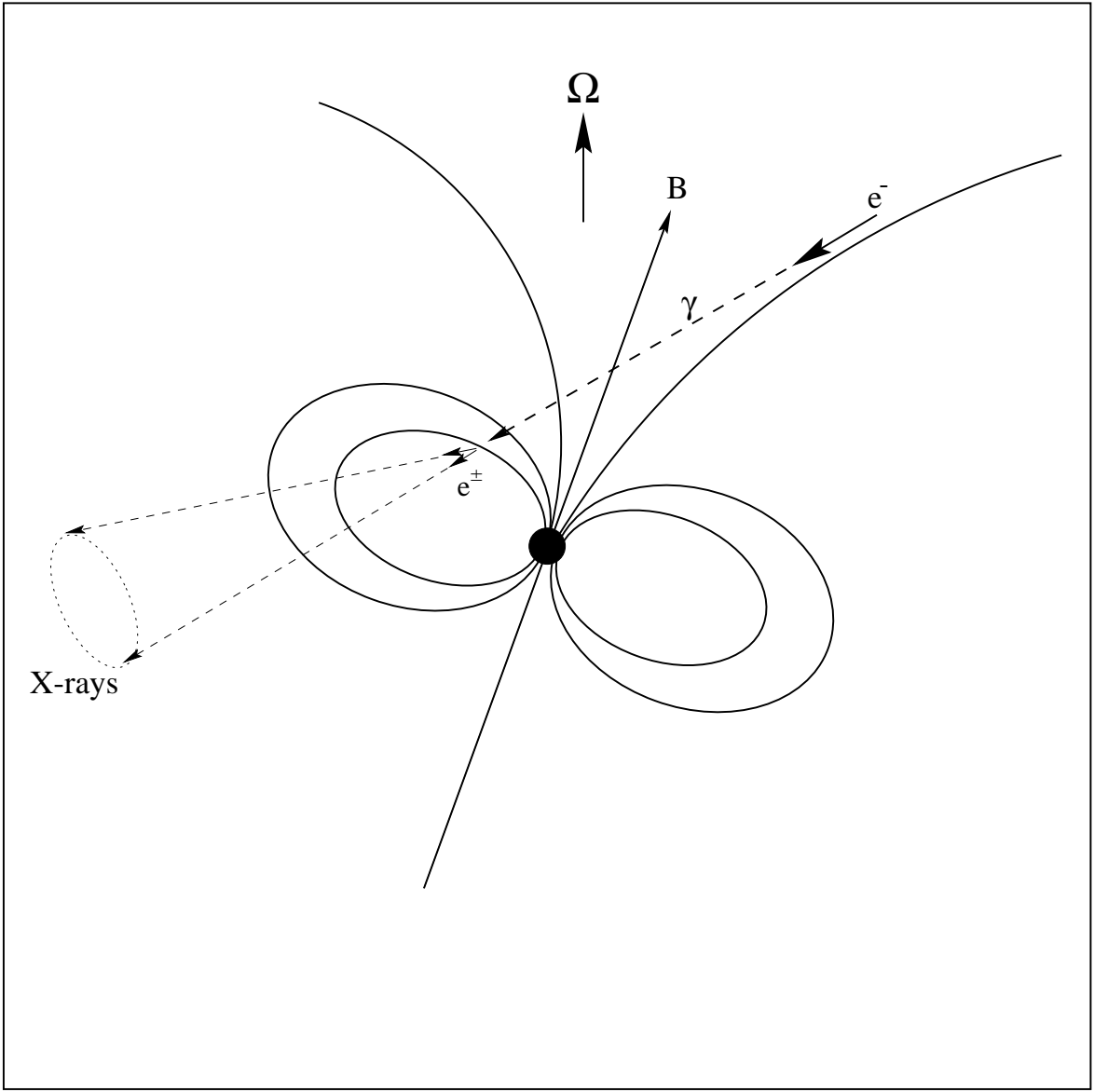}
\caption{Curvature \gr\ photons from inflowing primary
particles are converted to secondary \ep\ pairs when passing through 
closed field lines.  The \ep\ pairs will then synchrotron radiate a
power-law spectrum of energy index 0.5. \label{fig3}} 
\end{figure}

\placefigure{fig3}

The initial characteristic frequency of synchrotron radiation of these
locally created \ep\ is
\begin{equation}
\omega_{\rm c} \cong \gamma_{\perp}^{2} \gamma_{\parallel}\,\omega_{B}
\approx \gamma^{2} \sin \phi\,\omega_{B}.
\end{equation}
Their intensity spectrum is 
\begin{equation}
I_{\rm s}(\omega) \cong \frac{1}{2}\,\left(\frac{1}{\omega
\omega_{B} \sin \phi}\right)^{1/2}\,m c^{2}. \label{eq:power}
\end{equation}  
This power law spectrum with energy index 0.5 is valid from the minimum
value of the local cyclotron energy where the pairs are produced, up to
several MeV: 
\begin{equation}
E_{{\rm min}} \sim \frac{\hbar \omega_{B}}{\sin \phi} \sim 0.1 \ {\rm
keV},\label{eq:emin}
\end{equation} 
and
\begin{equation}
E_{{\rm max}} \sim \left(\frac{E_{\gamma}}{2m c^{2}}\right)^{2} \sin
\phi\,\hbar \omega_{B} \sim 5 \ {\rm MeV}. 
\end{equation} 
Because $\gamma$ in equation (\ref{eq:gamm}) is insensitive to details of
the initial accelerator, the energy index of the power law spectrum should
generally be $\cong 0.5$. This value agrees with the \asca\ spectra: both
Geminga and PSR 1055--52 are observed to have that energy index.  A
power-law spectrum with energy index 0.5 should be expected up to the MeV
range for all \gr\ pulsars.  The lower cutoff energy $E_{{\rm min}}$
depends on the local magnetic field in which the \gr\ is converted to \ep\
pairs and the pitch angle $\phi$. 

Both Geminga and PSR 1055--52 have been observed with the \hst;  Geminga
has a visual magnitude $V=24.5$ (\cite{big96a}), whereas PSR 1055--52 has
$m=24.88$ at $3400$ \AA\ (Mignani, Caraveo, and Bignami 1997). If the soft
X-ray spectrum is correctly described as a blackbody one, the visible flux
from Geminga is too large to be the long wavelength tail of the X-ray
spectrum (Halpern, Martin, \& Marshall 1996).  And in contrast to the
analysis by Mignani et al. (1997) of PSR 1055--52, we find the flux from a
blackbody extrapolation of the X-rays to be (2.0--3.1)  $\times 10^{-31} \
\rm ergs~cm^{-2}~s^{-1}~Hz^{-1}$, according to the best-fitted blackbody
temperature by Greiveldinger et al. (1996), much less than the observed
flux $1.3 \times 10^{-30} \ \rm ergs~cm^{-2}~s^{-1}~Hz^{-1}$ at 
$3400$ \AA .  While the extrapolated blackbody falls below the optical
flux, the fitted X-ray power-law component would exceed it.  This
observation conforms to our model, since we expect the low-energy cutoff
of the synchrotron spectrum to fall at about 0.1 keV.  Below the cutoff
energy the spectrum should rise (with increasing energy) at the canonical
low-frequency synchrotron index $-\frac{1}{3}$.  Within the uncertainty
associated with the power law best-fit, we estimate the cutoff energy to
be 20--500 eV. Improved X-ray, UV and optical observations are needed to
refine the cutoff energy. 

For a polar-cap accelerator, the crucial differences from the above are
the pitch angle $\phi$ and number of \gr\ photons $N_{\gamma}$ a primary
particle can make (to be discussed in \S 4.2).  For \gr s emitted above a
polar-cap accelerator in a dipole field, $\phi$ can never become much
greater than $(R \Omega / c)^{1/2}$, unless the emission radius greatly
exceeds $R$, in which case $B$ becomes too weak for pair creation.  If the
emission radius is much greater than $R$, equation (\ref{eq:paircvt}) will
not be satisfied, because $\phi$ increases proportionatedly to $r^{1/2}$,
while the magnetic field $B$ decreases as $r^{-3}$.  Those \gr s that are
converted would have $\phi \lesssim 10^{-1}$, and $E_{\rm min}$ will be
more than an order of magnitude greater than the spectral limit of
equation (\ref{eq:emin}).  A 1 keV lower bound does not seem to agree with
observations. 

\subsection{Luminosities}

The luminosity of the hard X-ray component (in \asca) from Geminga and PSR
1055--52 for an outer-magnetospheric accelerator model is
\begin{equation}
L_{\asca} \sim
f \dot{N}_{0} N_{\gamma} \int_{\omega_{1}}^{\omega_{2}}I_{\rm s}(\omega) 
\,d\omega \sim 10^{30} \  {\rm ergs \ s^{-1}}, \label{eq:lumi} 
\end{equation} 
with $\hbar \omega_{1} = 0.7 \ {\rm keV \ and} \ \hbar \omega_{2} = 5.0$
keV, the \asca\ coverage.  The factor $f$ in equation (\ref{eq:lumi}) is
the fraction of the Goldreich-Julian current that flows back out through
the accelerator to the polar cap. For an outer-magnetosphere accelerator
centered on the null surface ($\hat{\Omega} {\cdot \bf B} =0 $) this
fraction depends on details of the outer-magnetospheric accelerator model,
the angle between the spin and the dipole moment, etc.  It is typically
near $\frac{1}{2}$ for \gr\ pulsars not too far from their death lines
(CHRb).  The luminosity of equation (\ref{eq:lumi}) is close to the
observed values of $L_2$ in Table~1 for Geminga and PSR 1055--52.

For polar-cap models, only \gr s emitted by primary particles within a
stellar radius $R$ or so of the stellar surface can make \ep\ pairs (same
argument as \S 4.1).  Thus $N_{\gamma}$ will be reduced from the value in
equation (\ref{eq:multi}) by of order $R/s$, and the luminosity of
equation (\ref{eq:lumi}) should be reduced by this same factor.  The
simulations by Daugherty and Harding (1996) (Figure~3 in their paper) 
support our estimate. For \gr\ pulsars with $P \sim 10^{-1}$ s and a
dipole magnetic field approximation, $R/s \lesssim 10^{-2}$, so that
$L_{\asca} \lesssim 10^{28}$ \lum, not enough to account for what is
observed. 

\subsection{Light Curves}

If X-rays are produced above a polar-cap accelerator, \ep\ pairs have to
be created before $B$ drops below $\sim 10^{11}$ G.  Therefore the angular
size of the cone of emitted synchrotron radiation in X-rays should be of
order $(3R \Omega /c)^{1/2}$.  Our estimate is supported by the
simulations by Daugherty \& Harding (1996): their Figure 3 indicates that
pair production can no long be sustained if $r \gtrsim 3 R$.  The angular
width of the X-ray cone should then not much exceed the angular spread of
the open field line bundle from which the emission takes place.  Analyses
of radio emission generally place ``conal emission'' at $r \sim 10R$ and
``core emission'' at $r \sim R$ (\cite{ran90}, and references therein).
The synchrotron X-ray emission from these same field lines should then be
observed coincident in phase with the radio emission and with a width even
less than that of conal radio emission.  Neither is the case for PSR
1055--52, while no radio emission is yet confirmed from Geminga. 

\subsection{The Crab Pulsar}
 
The spectrum of the Crab pulsar in the \ro\ and \asca\ regime is dominated
by the radiation in two subpulses.  That radiation is almost certainly
from the same region and particles which are the source of its high-energy
(up to at least 1 GeV) \gr s, but not pairs created on closed field lines
in the near magnetosphere discussed above.  For X-rays from the latter we
should look at the radiation between these subpulses, which should also
have a broader angular spread than that in the two subpulses.  The
spectrum of this inter-pulse radiation has been isolated by Mineo et al.
(1997) (also see Massaro et al. 1998).  They find a power law for it, of
which energy index (smaller by 0.32 than that in the subpulses) is 0.5,
consistent with that for Geminga and PSR 1055--52.  This radiation extends
up to 200 keV. 

\section{Absence of Observable Hot Polar Caps}
 
Relativistic inflowing particles radiate away much of their energy before
reaching the polar cap. According to equation (\ref{eq:res})  the residual
energy of the charged particles impacting the polar cap is
\begin{equation}
E_{\rm f}=mc^{2} \left[\frac{2\Omega e^{2}}{mc^{3}}\ln\left(\frac{r_{\rm
LC}}{R}\right)\right]^{-1/3} \sim 5.5 \ {\rm ergs}.  \label{eq:rener}
\end{equation}
With constant bombardment of these particles at a rate of $10^{32} \ \rm
s^{-1}$ (the Goldreich-Julian current) the polar cap would radiate
X-rays with luminosity
\begin{equation}
L_{\rm X} = f E_{\rm f} \dot{N}_{0} \cong 2.1
\times 10^{32} \ {\rm ergs \ s^{-1}}. \label{eq:lumx}
\end{equation}
However, no evidence for such a strongly heated polar cap is found in
Geminga or PSR 1055--52. The total thermal X-ray luminosity of PSR
1055--52 is $2.3 \times 10^{32}$ \lum, whereas for Geminga it is $1.5
\times 10^{31}$ \lum.  In both cases X-rays seem to be emitted from a
large fraction of the surface of the neutron star.  We attribute the
absence of observed hot polar caps to the formation of a reflective \ep\
``blanket'' above the stellar surface. 

The creation of $10^{37}$ \ep\ pairs per second which produce the
power-law synchrotron radiation in \S 4 would take place mainly on closed
field lines in which $B \sim 10^{10}$ G (at $r \sim (\frac{B_{\rm
p}}{10^{10}})^{1/3} R \sim 6R$). As we discuss later, production of \ep\
pairs at such a rate would have significant impact on the observability of
X-ray emission from the stellar surface.  Because we expect a \gr\ pulsar
with an outer-magnetospheric accelerator to produce this number of \ep\
pairs, but not so for an ordinary pulsar, the X-ray spectrum is expected
to be different for these two families. 

In order for a blanket to be sustainable, the \ep\ annihilation rate in it
must not exceed the pair injection rate there.  This annihilation rate
for a local $n_{\pm}$ is 
\begin{equation}
r_{\pm} \sim (n_{\pm})^{2}\,\sigma_{\pm}\,\left(\frac{c}{v_{\pm}}\right)
\,v_{\pm} \sim
10^{14} (n_{\pm})^{2}_{14} \ \rm cm^{-3} \ s^{-1},
\end{equation}
in which $\sigma_{\pm} \frac{c}{v_{\pm}}$ is the annihilation cross
section for non-relativistic \ep\ pairs with velocities $v_{\pm}$.  It
corresponds to a total near magnetospheric annihilation rate
\begin{equation} R_{\pm} \sim 10^{34}\,(n_{\pm})^{2}_{14} \ \rm s^{-1}.
\label{eq:rate} \end{equation}

In comparison the \ep\ production rate on the closed field lines from
equations (\ref{eq:GJ}) and (\ref{eq:multi}), $N_{\gamma} \dot{N_{0}} \sim
10^{37} \ \rm s^{-1}$, could balance $R_{\pm}$ with $n_{\pm} \sim 5 \times
10^{15} \ \rm cm^{-3}$.  When a blanket is formed the lifetime of an \ep\
pair is limited by the pair annihilation rate to about one second when the
pair density is $10^{14} \ \rm cm^{-3}$.  If no blanket were to form and
positrons flowed without interruption to the stellar surface, their
residence time in the near magnetosphere would be a thousand times less. 
In that case there would not be a sufficient pair density to surround the
star almost everywhere with a good reflecting blanket. 

There are five forces acting on these \ep\ pairs:

(i) the gravitational force,
\begin{equation}
{\bf F}_{\rm g} = - \frac{GMm}{r^{2}} \hat{\bf r}.
\end{equation}

(ii) the centrifugal force,
\begin{equation}
{\bf F}_{\rm c} = m \Omega^{2} \bf r.
\end{equation}

(iii) an electric force, which may exist within the open field line bundle
\begin{equation}
{\bf F}_{\rm E} = e (\bf E \cdot \hat{B}) \hat{\bf B}.
\end{equation}

(iv) the Lorentz force which confines \ep\ to the local magnetic field
lines, and

(v) the radiation force of scattered X-rays originally emitted  
from the surface of the neutron star.

We specialize our numerical calculations below to PSR 1055--52, but the
formulae are applicable to other pulsars.  The strong magnetic field
restricts particle motion across a field line, therefore we consider only
the component of forces along the magnetic field.  Because ${\bf E \cdot
B}=0$ on closed field lines, and centrifugal force is insignificant in the
region of interest [$r \sim$ 2--3 $R$, to be shown later in equation
(\ref{eq:rpeak})], it is sufficient to consider merely gravity and
radiation.  Zhu and Ruderman (1997), in a study of the \ep\ annihilation
line from the Crab pulsar, found that a balance between these forces can
result in a stable equilibrium, in which an optically thick accumulation
of \ep\ pairs is formed over the neutron star.

A blackbody spectrum for polar-cap emission from the surface is 
\begin{equation}
I_{\rm b}(\omega)= L_{\rm X}\,\frac{15}{\pi^{4}}
\,\frac{\hbar^{4}}{(kT)^{4}}
\,\frac{\omega^{3}}{\exp(\hbar \omega/kT)-1},
\end{equation}
with $L_{\rm X}$ the bolometric luminosity ($2.1 \times 10^{32}$ \lum), 
and $T_{2}$ the polar cap temperature which is estimated to be 
\begin{equation}
kT_{2} \sim \left(\frac{L_{\rm X}}{\pi R^{3} \frac{\Omega}{c}
\sigma_{\rm SB}}\right)^{1/4} 
\sim 0.5 \ {\rm keV},
\end{equation}
in which $\sigma_{\rm SB}$ is the Stefan-Boltzmann constant.  Here we
assume the canonical polar cap size.  Our purpose is only to estimate but
not obtain a definite numerical result, and later we will find out that
the radial position of the blanket ($r_{\rm pk}$) is insensitive to
temperature [$r_{\rm pk} \propto T^{-1/3}$, see equation
(\ref{eq:rpeak})]. 

The magnetic field for a dipole at the center of the neutron star is
\begin{equation}
{\bf B}(r,\theta)=\frac{B_{\rm p}}{2}\,\left(\frac{R}{r}\right)^{3}
\,\left(2 \cos \theta\,\hat{\bf e}_{\rm r} + \sin \theta\,\hat{\bf e}_{\theta}
\right).
\end{equation}
In a strong magnetic field the averaged $e^{-}$-$e^{+}$ cross section 
for X-ray scattering is approximated by (\cite{bla76})
\begin{equation}
\sigma = \sigma_{\rm T}\,(\hat{\epsilon} \cdot \hat{\bf B})^{2} + 
\frac{2 \pi^{2}e^{2}}{mc} \mid \hat{\epsilon} \times \hat{\bf B} \mid^{2}
\delta(\omega_{B}-\omega), \label{eq:cross}
\end{equation}
with $\hat{\epsilon}$ the photon polarization and $\sigma_{\rm T}$ the
Thomson cross section. In the optically thin case, the radiation force
averaged over photon polarization is
\begin{equation}
{\bf F}_{\rm r} = \left(\frac{\pi^{2} e^{2}}{4 m c^{2}}\right)
\,\left(\frac{1+7 \cos^{2}\theta}
{1+3 \cos^{2} \theta}\,\frac{I_{\rm b}(\omega_{B})}{r^{2}}\right) \hat{\bf r}.
\label{eq:radf}
\end{equation}

The angle-averaged radiation force is plotted in Figure 4a.  For the
estimated polar cap temperature 0.5 keV, the stable equilibrium point is $
r=6.9 R$.  Electrons and positrons produced below this point will be
pushed out along their magnetic field line to this radius; those above it
will be pulled in toward it by gravity. When the number density of
electrons and positrons grows large enough, the \ep\ plasma becomes
optically thick to the X-ray photons from the surface and the radiation
force per lepton of equation (\ref{eq:radf}) decreases. To make an
optically thick layer of \ep\ pairs for the polar cap X-rays, the required
minimum number density is
\begin{equation}
n_{\rm t}(r,\theta) \cong \frac{3}{2} 
\,\frac{B_{\rm p}}{\pi^{2} e R}\,\left(\frac{R}{r}\right)^{4}\,
\frac{(1+3\cos^{2}\theta)^{3/2}}{1+7\cos^{2}\theta}.
\end{equation} 
A typical $n_{\rm t}$ at $r=2R$ is $\sim 5 \times 10^{13} \ \rm cm^{-3}$. 
The maximum pair density that can be supported by the X-ray radiation
force is reached when the total radiation force from complete absorption
equals the gravitational attraction. An optically thick \ep\ blanket
begins to develop inward from the stable equilibrium point $r = 6.9 R$ but
cannot grow to smaller radii than the unstable equilibrium point $r=1.3
R$.  The radiation force for this optically thick example is plotted in
Figure 4b: between $1.3 R$ and $6.9 R$, the radiation force equals the
gravitational force.  Below the unstable equilibrium point \ep\ pairs are
pulled by gravity back to the surface of the neutron star. The maximum
supportable density is achieved when all of the initial momenta of the
resonance scattered X-ray photons support the \ep\ plasma, then
\begin{equation}
n_{\rm max} (r,\theta) \cong
\frac{45}{4 \pi^{5}}\,\frac{1}{G M m R c}
\,\left(\frac{\hbar e B_{*}}{m c k T}\right)^{4}
\,\left(\frac{R}{r}\right)^{13} \frac{L_{\rm X}}
{\exp[\frac{\hbar e B_{*}}{m c k T} (\frac{R}{r})^{3}]-1}, \label{eq:nmax}
\end{equation}
in which $B_{*}=\frac{B_{\rm p}}{2}(1+3 \cos^{2} \theta)^{1/2}$. 
The number density $n_{\rm max}$ is plotted in Figure 5; its maximum
occurs at
\begin{equation}
r_{\rm pk} \cong 0.616 \,R\,\left(\frac{\hbar e B_{\rm p}}{m c kT}
\right)^{1/3}.\label{eq:rpeak}
\end{equation}
If the local density of \ep\ pairs is greater than the maximum supportable
value, pairs will be pulled toward to the star by gravity, otherwise they
will be repelled from the star by radiation.  When a steady state is
reached, the local number density will be near the maximum supportable
value, but essentially zero elsewhere because it drops sharply after the
resonant peak [equation (\ref{eq:nmax})]. The above formulae are valid on
closed field lines, as mentioned earlier.  However, the open field bundle
subtends only a small solid angle, $ \pi (2 R \Omega /c) \sim 6 \times
10^{-3}$ ster, thus the star is well blanketed. 

\begin{figure}
\plottwo{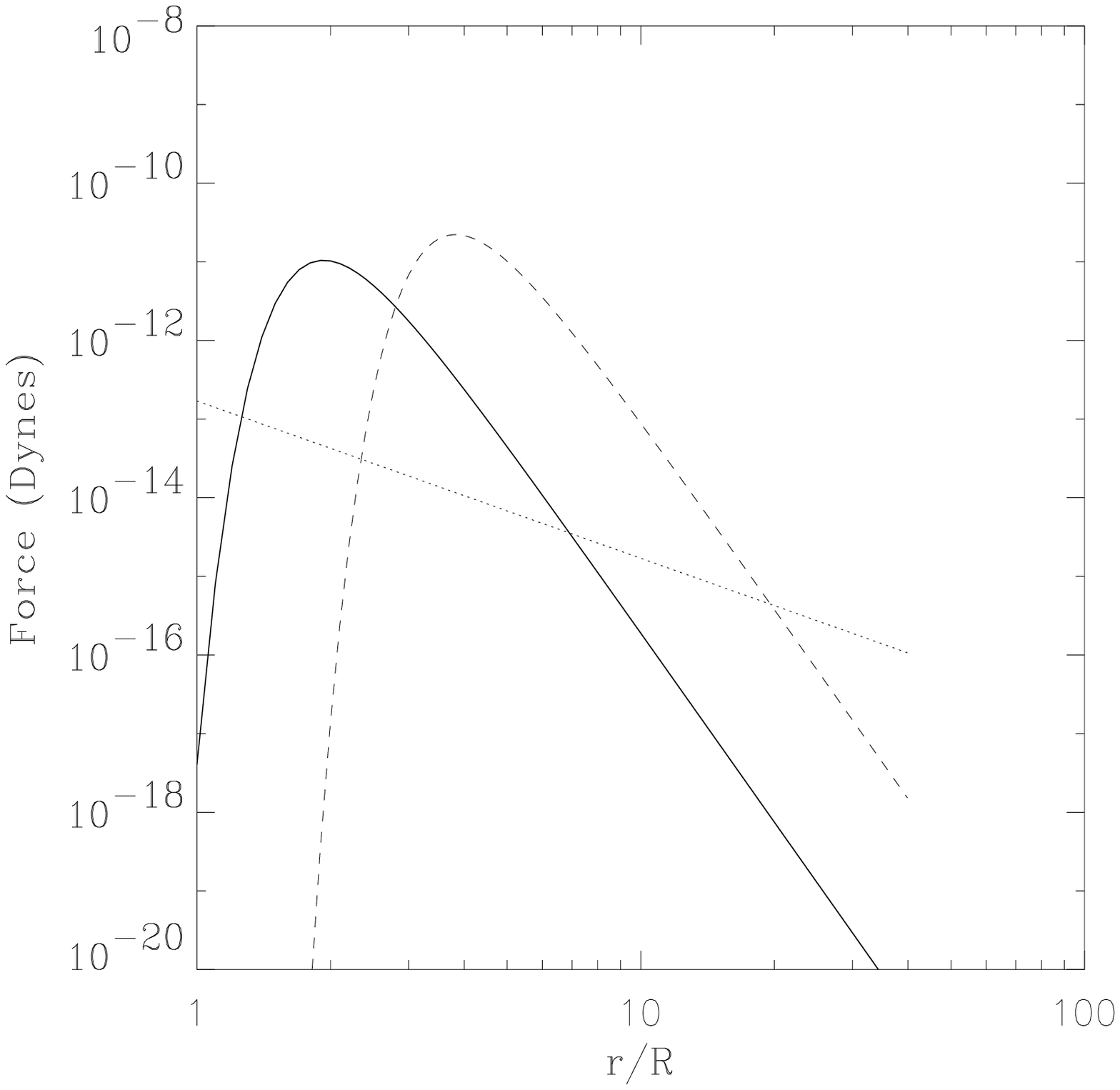}{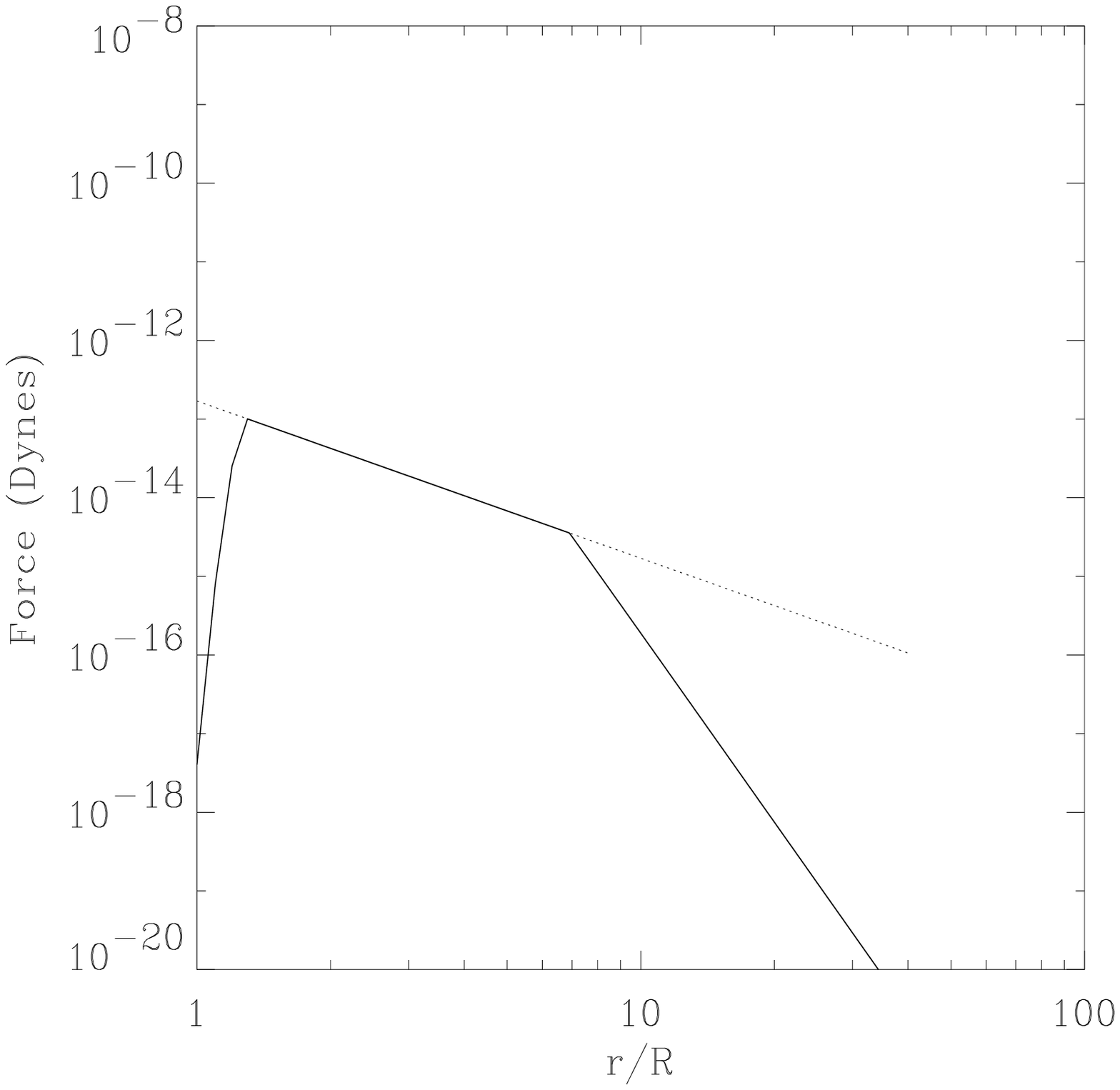}
\caption{(a) The optically thin case for the radiation force and the
gravitational force on \ep\ as a function of distance $r$ from the center
of the neutron star; the solid line indicates the radiation force for $kT
= 0.5$ keV, the dashed line for $kT = 0.06$ keV.  The dotted line gives
the gravitational force.  (b) The optically thick case for the radiation
force and the gravitational force; these two forces are equal between the
two equilibrium points. 
\label{fig4}}
\end{figure}

\begin{figure}
\plotone{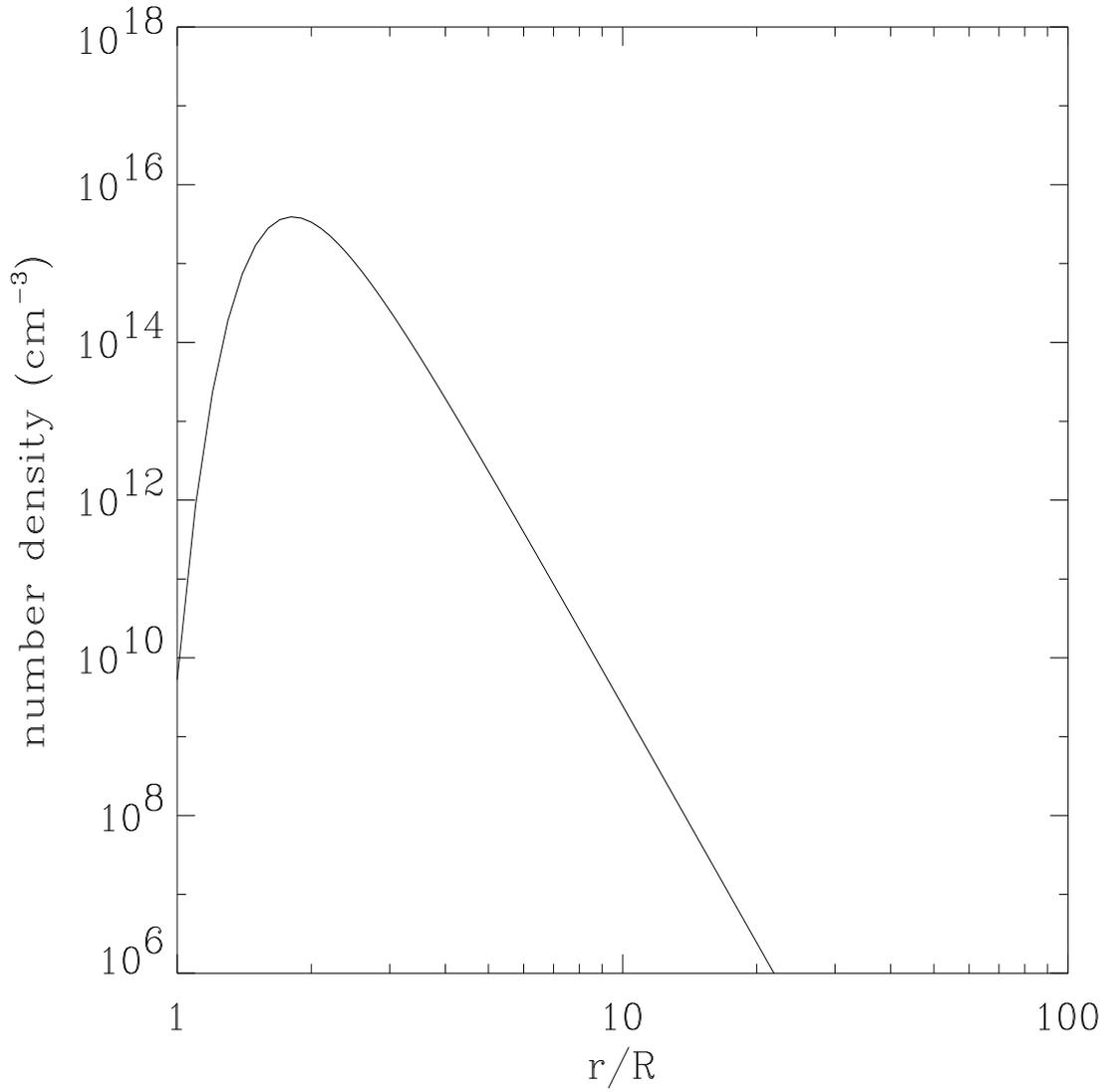}
\caption{The maximum number density supportable by the
radiation force as a function of distance $r$ for $kT=0.5$ keV and 
$L_{\rm X}=2.1 \times 10^{32}$ \lum. \label{fig5}} 
\end{figure}

\placefigure{fig4}
\placefigure{fig5}

With $kT_{2} \cong$ 0.5 keV, the blanket has maximum number density
$n_{\rm max} \cong 5 \times 10^{15} \ \rm cm^{-3}$ at radius $r_{\rm pk}
\cong 2 R$.  This is almost three orders of magnitude greater than that
required for an optically thick blanket. This optically thick blanket
prevents direct observation of the polar cap from most directions, because
the unblanketed part only occupies $\sim 10^{-3}$ the total area of the
sphere of radius $2R$. Most of the photons from the polar cap will be
reflected back to the surface of the neutron star (Figure~6).  This
process transfers emitted polar cap X-ray energy to the entire surface of
the neutron star, from which it is reradiated away at a lower temperature
\begin{equation} 
T_{1} \sim \left(\frac{L_{\rm X}}{4 \pi R^{2}
\sigma_{\rm SB}}\right)^{1/4} \sim 7.4 \times 10^{5} \ \rm K. 
\end{equation} 
This temperature is in reasonable agreement with the observation.     
The contribution of cyclotron resonant back-scattering onto the stellar   
surface gives rise to thermal radiation from the surface which is in
addition to that from the cooling of the neutron star which was presumably
very hot when it was formed; it might explain the huge difference of soft 
X-ray luminosities for Geminga and PSR 1055--52 discussed below.  The
cooler surface radiation will have to pass through its own blanket at $r  
\sim 5 R$ or through the open field line ``holes'' where \ep\ blankets can
be kept from forming by the strong relativistic current flows. We discuss
the escape of soft X-rays further in \S 8. 

\section{Resonant Inverse Compton Scattering on Radio Photons}

If the true ages of Geminga and PSR 1055--52 are near their characteristic
ages, it is difficult for a cooling model to explain their enormous
difference in luminosity since they have similar characteristic ages, and
Geminga is younger. One possible explanation is illumination of the
extremely relativistic inflowing particles by even a tiny fraction of the
radio emission which is generally assumed to be generated by a polar cap
accelerator.  Although no such radio-frequency radiation has been observed
from Geminga, it is not implausible that this is because we do not
intercept its emission beam. Special near-surface magnetic field
configuration (especially for core radio beam emission) or gravitational
bending of \gr s might allow interception of a small part of the radio
beam far from the stellar surface by the postulated extreme relativistic
inflow of particles directed down toward the polar cap.  Then cyclotron
resonant inverse Compton scattering between these relativistic
$e^{-}$($e^{+}$) and the radio beam may become important.  In the rest
frame of the inflowing particles, the radio frequency is upshifted by
$\gamma$ ($\omega' \approx \gamma \omega$).  For a radio photon of
frequency 500 MHz and $\gamma \sim 10^{7}$, there is a resonance at $B
\approx 10^{9}$ G or $r \approx 10 R$.  When inverse Compton scattering
occurs, the final scattered photon energy is
\begin{equation} 
E_{\gamma} \sim \gamma^{2} \hbar \omega. 
\end{equation} 
For $\gamma \sim 10^{7}$, and frequency of radio photons $\sim 500$ MHz,
the final energy of these scattered photons is $\sim 100$ MeV, and they
will be converted to \ep\ pairs before approaching the surface.  The \gr\
energy will thus be converted into the synchrotron radiation as discussed
in \S 4.   

With the resonant cross section $\sigma_{\rm r} \cong \frac{2
\pi^{2}e^{2}}{m c} \delta(\omega_{B}-\omega)$, and a Goldreich-Julian
number density for inflowing primaries, the optical depth for the radio
photons passing through the inflowing particles is
\begin{equation}
\tau_{\rm i} = \int \sigma_{\rm r} n dr \cong
\frac{\pi}{3} \frac{r_{\rm i} \Omega}{c} \cong 10^{-2},
\end{equation}
with $r_{\rm i}$ the resonance radius $\sim 10 R$.  Hence only a very
small fraction of any incident number of radio photons would be scattered. 
[Because the fraction of radio photons interacting with leptons is
negligible, inverse Compton scattering cannot be responsible for the
absence of radio emission from Geminga.] The energy loss rate of charged
particles to resonant inverse Compton scattering would then be
\begin{equation}
\dot{E_{\rm e}} \sim
\tau_{\rm i} \cdot \frac{f_{\rm r} L_{\rm radio}}{\hbar \omega}
\cdot \gamma^{2} \hbar \omega
\sim 10^{38} f_{\rm r} \ \rm erg \ s^{-1},
\end{equation}
with $f_{\rm r}$ the small fraction of the radio beam that encounters
relativistic particles and $L_{\rm radio} \sim 10^{25}$ \lum\ the radio
luminosity. Although this estimate involves many uncertainties, it shows
that resonant inverse Compton scattering can be extremely efficient in
transferring energy from particles into \gr s: a very minuscule fraction
$f_{\rm r} \gtrsim 10^{-6}$ could take away most of the energy of the
primary inflowing particles and reduce polar cap heating.  Such a
mechanism would not necessarily be a universal one and there is no obvious
reason that polar-cap inflow in PSR 1055--52 must be similarly affected. 

\section{Heated Polar Cap Areas} 

The backflow of extremely relativistic particles heats the polar cap
region.  The predicted polar cap heating varies among models
(\cite{che80}; \cite{aro81a}; \cite{hal93}). The residual energy $E_{\rm
f}$ of each relativistic particle is about 5 ergs [equation
(\ref{eq:rener})] from an outer-magnetospheric accelerator whose inflow is
uninterrupted (i.e., not the case proposed for Geminga in \S 6), whereas
$E_{\rm f} \sim e \Delta V \sim 1$ erg for a polar-cap accelerator.  An
outer-magnetospheric accelerator can be a copious supplier of charged
particles; nearly the full Goldreich-Julian current would be expected to
pass through it.  The full Goldreich-Julian current might not be achieved
for backflow through a polar cap accelerator and we can estimate only an
upper limit for the heated polar-cap luminosity in that case
(\cite{aro81b}).  As discussed in \S 5, a polar cap heated by an
outer-magnetospheric accelerator is not expected to be directly visible
unless an observer looks in along the dipole axis. We consider below only
the observable polar cap heating of ordinary pulsars.  The three pulsars
in Table 1 with thermal emission from a tiny area are prospective
candidates for such polar cap heating.  However, their temperatures and
luminosities vary greatly.  While PSR 1929+10 and PSR 0950+08 have $\sim
10^{-2}$ smaller polar cap areas than that of central dipole models, PSR
0656+14 has a 30 times larger one (Table 1). We emphasize that the fitted
luminosity and area depend on the assumed viewing geometry and distance,
as well as on possibly important atmosphere effects (e.g. Meyer, Pavlov,
\& M\'{e}sz\'{a}ros 1994).

The total magnetic flux $\Phi$ out of a polar cap is determined by the
observed pulsar spin-down power ($I \Omega \dot{\Omega}$):
\begin{equation}
A_{\rm pc} B_{\rm pc} \equiv \Phi \sim \left(\frac{c I
\dot{\Omega}}{\Omega}\right)^{1/2}.
\end{equation}
For a central dipole model the polar cap area 
\begin{equation}
A_{\rm pc} = A_{0} \equiv \pi R^{3}\,\frac{\Omega}{c},
\end{equation}
and
\begin{equation}
B_{\rm pc}=B_{\rm p}.
\end{equation}

An anomalously large (small) polar cap area $A_{\rm pc}$, relative to
$A_{0}$, corresponds to a polar cap $B_{\rm pc}$ which is much smaller
(larger) than $B_{\rm p}$.  While a central dipole model gives a good
description for the magnetic field far from the star, it may be inadequate
for the surface field.  For example, an off-center dipole, special
evolutionary features not describable by a simple dipole, or effects of
the strong diamagnetism of the superconducting core of the star may all
lead to a $B_{\rm pc}$ very different from $B_{\rm p}$.

We consider below two possible consequences for $B_{\rm pc}$ and $B_{\rm
p}$ which follow from an evolutionary model for the surface field of
spin-down pulsars.

\subsection{PSR 0656+14}

PSR 0656+14 shares many similarities with Geminga and PSR 1055-52, but
detection of \gr\ emission from it remains weak and inconclusive
(\cite{ram96}). PSR 0656+14 lies very close to the \gr\ death line, which
makes it difficult to decide theoretically if it should be a \gr\ pulsar.
According to the model in \S 5, there is an important connection between
strong \gr\ emission and the nature of the hard X-ray component in the
sense that a power law of energy index 0.5 is to be associated with a \gr\
pulsar.  In addition, an \ep\ resonance blanket is produced by a \gr\
pulsar, which should prevent us from observing the heated polar cap
directly.  Since the X-ray spectrum of PSR 0656+14 can be fitted with a
double blackbody model (Table~1), we infer that it should not be a \gr\
pulsar, or at least not one with the magnitude of particle inflow of an
outer-magnetospheric accelerator. From the observed temperature and
luminosity of the hotter component, the polar cap size appears to be about
30 times greater than the canonical value $A_{0}$.  A polar cap with
lower temperature and larger area implies a locally much weaker magnetic
field than that of a central dipole.  We attribute this difference to the
circumstance that PSR 0656+14 is a nearly aligned pulsar (\cite{lyn88}).

The interior of a neutron star consists of superfluid neutrons, and
superconducting protons.  A spinning neutron star with an angular velocity
$\Omega$ has an array of quantized vortex lines parallel to its spin axis.
Any magnetic field passing through the superconducting protons of the core
is expected to be organized into quantized flux tubes below the stellar
crust.  In contrast to the quasi-parallel neutron vortex line array, the
flux tube array is expected to have a complicated twisted toroidal and
poloidal structure to achieve magnetohydrodynamic stability in the deep
interior of a conducting star (\cite{flo77}). The vortex array of a
spinning-down neutron star must expand with stellar spin-down. Because the
core of a neutron vortex and a flux tube interact strongly as they pass
through each other, the moving vortices push or pull on the flux tube
array.  This interaction results in flux tubes either moving together with
the neutron vortices, or being cut through if the flux tube array cannot
respond fast enough.  The superfluid vortices move outward with a velocity
\begin{equation}
v_{\rm V}=r_{\perp} \frac{\dot{P}}{2 P},
\end{equation}
with $r_{\perp}$ the distance from the vortex to the spin axis.  Near the
spin axis, the vortices move slowly; away from it, the vortices move much
faster.  If we assume that a pulsar was created with an almost
axisymmetric magnetic field on its surface, and the axis of magnetic
dipole happens to be almost aligned with the spin axis (the apparent case
for PSR 0656+14), the fast moving outer vortices cut through the magnetic
flux tubes, while much more slowly moving inner vortices drag the magnetic
flux tubes with them and leave behind a region with a much weaker magnetic
field around the spin axis (Figure 7).  The last closed field lines become
very distorted near the surface, and the actual polar cap will be larger
than that of a pure central dipole field.  [According to the model of
Arons and co-workers (Scharlemann, Arons, \& Fawley 1978), polar cap
accelerators form only on ``favorably'' curved field lines, those bend
toward the axis of rotation.  An aligned pulsar with a central dipole
field would have no favorably curved lines, unlike the configuration of
Figure 7, which would give rise to strong \ep\ production above the polar
cap.] If the pulsar were born with period $P_{\rm i} \sim 20$ ms, the
present magnetic field near the spin axis should be diminished by a factor
$P/P_{\rm i} \sim 20$, and the region close to the spin axis should have a
weak magnetic field $B_{\rm pc} \sim 10^{11}$ G.  The expected polar cap
temperature should accordingly be lower than in a central dipole model:
\begin{equation}
T \sim \left(E_{\rm f} \frac{\Omega B_{\rm pc}}{2 \pi e}
\frac{1}{\sigma_{\rm SB}}\right)^{1/4}
\sim 2 \times 10^{6} \ \rm K.
\end{equation}

\placefigure{fig7}

We emphasize that it is only when an important platelet, one that contains
the polar cap, happens to be pierced by the spin-axis that it can be
pulled apart in this way during spin down.  This is an exception to the
usual case in which the platelets remain intact (\S 7.2).  We also note
that for this special geometry the total dipole field is least affected by
the spin-down history of the star.  The original field has been moved away
from the spin axis but has not been pushed out of the core of the star.
[Despite its age $\tau \sim 1.1 \times 10^{5}$ years, PSR 0656+14 has an
average surface dipole field $\approx 4.7 \times 10^{12}$ G, very close to
that of a young pulsar.] It is only after such expulsion has occurred that
reconnection can begin to diminish greatly the dipole moment of the
spinning pulsar (Ruderman, Zhu, \& Chen 1998). 

\subsection{PSR 1929+10 and PSR 0950+08}

PSR 1929+10 and PSR 0950+08 are two relatively old pulsars.  Their
surfaces are expected to have cooled to less than $10^{5}$ K
(\cite{nom87}; \cite{lat91}; \cite{pag92}).  Therefore, the X-rays from
these pulsars detected by \ro\ and \asca\ and fitted by temperatures $ > 1
\times 10^{6}$~K, cannot be coming from the cooling surface but rather
from small areas that are reasonably associated with polar cap heating.

Because of their small spin-down power ($\sim 10^{33}$ \lum), both pulsars
are probably unable to sustain outer-magnetospheric accelerators (e.g.
\cite{ch93a}). Both pulsars have rather weak surface dipole magnetic
fields, $B_{\rm p} \approx $(2--5)$ \times 10^{11}$ G.  This might be an
indication of an evolution of the magnetic field from an initially
stronger one. As a young pulsar spins down, the initial hot polar caps
move with the highly conducting crust in which they are embedded.  When
$P$ reaches a critical value $P_{\rm c}$, substantial cracking of cool
crust begins.  In the subsequent spin-down, when $P > P_{\rm c}$, the
average surface dipole field $\langle B(P)\rangle$ decreases inversely
with $P$ because the movement of the core superfluid neutron vortex array
pulls core flux and the crust through which it penetrates (\cite{rud91}; 
Chen, Ruderman, \& Zhu 1998), so
\begin{equation}
\langle B(P)\rangle = \langle B(P_{\rm c})\rangle P_{\rm c}/P.
\end{equation}
Relatively strong uniform platelet fields $(B_{\rm pl})$ remain frozen at
\begin{equation}
B_{\rm pl} = \langle B(P_{\rm c}) \rangle \sim B({\rm Vela}) \sim 3 \times
10^{12} \ \rm G.
\end{equation}
these platelets move away from each other during the spin-down era but the
stress within them is less able to break them apart (except in the special
case of \S 7.1). The more dilute surface flux is then concentrated in
small areas.  The area of a polar cap is
\begin{equation}
A_{\rm pc} \sim \frac{\langle B(P)\rangle}{B_{\rm pl}} \cdot A_{0} 
\sim \frac{1}{10} A_{0}. 
\end{equation} 
The temperature of the heated part of the polar cap is
\begin{equation} 
T \sim \left(E_{\rm f} \frac{\Omega
B_{\rm pl}}{2 \pi e} \frac{1}{\sigma_{\rm SB}}\right)^{1/4} \sim 5 \times 10^{6}
\ \rm K.
\end{equation} 
This result agrees with observation, $T \sim 5 \times 10^{6}$ K for PSR
1929+10 and $T \sim 6 \times 10^{6}$ K for PSR 0950+08.

\section{Soft X-ray Modulation}

Three of the pulsars discussed in this paper are detected in soft X-rays,
presumably thermal emission from the entire surface of the neutron stars.
The emission is highly modulated, with a typical pulsed fraction of 20\%.
This feature implies an anisotropic structure on or near the surface of a
neutron star.  The blanket model of \S 5 gives such an anisotropy.  For
the \gr\ pulsars Geminga and PSR 1055--52, copious \ep\ pair production on
closed field lines is a plausible consequence of strong \gr\ production. 
PSR 0656+14, which has a weak field polar cap region surrounded by a
stronger field closed field line region (\S 7.1), should have copious \ep\
pair production on closed field lines surrounding its polar cap
accelerator.  A cyclotron resonant blanket could form around the hot polar
cap.  Thus, we expect an optically thick surrounding blanket above all
these three neutron stars.  X-ray photons will escape blanketed neutron
stars either through the two holes of the blanket on their open field
lines, or by diffusing through the blanket itself.  The ratio of these two
fluxes depends on the size of the holes relative to the area of the
blanket, and the probability of penetrating through the blanket. If a
photon falls on the hole, the probability of penetration is one; if a
photon hits the blanket, the probability of penetration is proportional to
$(1+\tau_{\rm b})^{-1}$ in which $\tau_{\rm b}(\omega)$ is the
frequency-dependent optical depth. [This should include both diffusion
through the blanket and the special incident angles and polarizations for
which $\hat{\epsilon} \times \hat{\bf B} =0$ is a good approximation in
equation (\ref{eq:cross}).] Then fraction the flux through the holes
($f_{\rm h}$)  versus that through the blanket ($f_{\rm b}$) is
\begin{equation} \frac{f_{\rm h}}{f_{\rm b}} = \frac{2 \pi r^{3}
\frac{\Omega}{c}}{4 \pi r^{2}} (\tau_{\rm b}+1). \end{equation} The
blanket area is roughly $10^{3}$ times greater than the hole size, and the
expected optical depth $\tau_{\rm b}$ is roughly 10--200, depending on the
photon frequency, incident angle, and polarization.  As the star spins the
photons escaping from the blanket should be more uniformly distributed
than those from the holes.  Then the strongly modulated portion of the
light curve of soft X-rays could be from hole emission and the relatively
non-varying part from passage through the blanket. 

In fitting theoretical spectra to observed ones it should be emphasized
that the areas and attitudes of both the open field holes in the blanket
through which ``hohlraum'' radiation is mainly emitted and the blanket
transmission $(1+\tau_{\rm b})^{-1}$ are energy dependent.  When $f_{\rm
b} \ll f_{\rm h}$ and $r \gg (c R^{2}/\Omega)^{1/3}$, the emitted
radiation resembles that from a classical ``hohlraum'' at the temperature
of the stellar surface.  Radiation is almost all reflected by the blanket,
and absorbed and reemitted by the stellar surface before it escapes
through the two small open field line holes.  X-rays emerging from the two
open field line holes would not be observed to have the canonical
blackbody spectrum because the ``hohlraum'' radiation would have an
emitting area which depends on the photon energy $E$,
\begin{equation}
A_{\rm h}=\left(\frac{\pi R^{3} \Omega}{c}\right) \left(\frac{\hbar e
B_{\rm p}}{mcE}\right).
\end{equation}
This energy-dependent area is independent of the surface field structure
when local $B \ll B_{\rm p}$, so $A_{\rm h}$ is not an adjustable
parameter.  However, quantitative comparisons of observations and models
are not yet available because of uncertainties associated with luminosity
measurements and the density distribution of blanket \ep\ pairs needed to
determine $\tau_{\rm b}$.

\section{Conclusions and Future Observations}

We have argued that {\it if} the keV X-ray emission from \gr\ pulsars is
indeed synchrotron radiation with an energy index 0.5, then
outer-magnetospheric accelerators are consistent with the energy range
over which this power law is observed, the X-ray luminosity in the
observed 0.7--5.0 keV range, and the broad modulated light curve. We find
it difficult to see how polar-cap accelerator models for this synchrotron
radiation can give these same results.  If the expected $10^{37} \ \rm
s^{-1}$ \ep\ pairs are produced on the closed field lines above the
stellar surface of \gr\ pulsars, cyclotron resonant blankets are formed
which greatly obscure X-rays emitted from that surface. Observations are
not inconsistent with what is expected in the presence of such blankets.

Our model implies that a \gr\ pulsar should have a power-law spectrum of
energy index 0.5 in X-rays extending from 0.1 keV to a few MeV, but polar
cap heating will not usually be directly observable.  This can be tested
with more sensitive higher-energy X-ray data on the pulsars discussed in
this paper, as well as others.  These data can be obtained with longer
exposures by \asca, or by the next generation of X-ray missions such as
{\it AXAF}, {\it XMM}, and {\it ASTRO-E}. The validity of the power-law
interpretation of the hard X-ray flux should also be tested by observing
its extension to a few MeV, and its low-energy cutoff at $\sim 0.1$~keV
with improved soft X-ray, UV, and optical spectra. Currently \ro\ data
show that, in addition to the \gr\ pulsars listed in Table 1, PSR 1951+32
also has a power-law spectrum of energy index $0.6^{+0.2}_{-0.2}$
(Safi-Harb, \"{O}gelman, \& Finley 1995); this result is an encouraging
compliment to the model.  Additional data from \asca\ should be extremely
relevant. 

For \gr\ pulsars, creation of $10^{37} \ \rm s^{-1}$ \ep\ pairs is
expected.  There would be an expected $e^{+} + e^{-} \rightarrow \gamma +
\gamma$ slightly red-shifted annihilation line of twice this strength.  In
the case of the Crab pulsar and possibly some other \gr\ pulsars,
mechanisms exist which might raise this estimate by 10--$10^{2}$ [e.g. 
some of the few MeV \gr s become converted at $B \sim 10^{12}$ G, see Zhu
and Ruderman (1997)].  A line this weak is not detectable (e.g. by OSSE)
at typical \gr\ pulsar distances.  It might be observed if a \gr\ pulsar
is someday found at a distance $\sim 100$ pc or less. 

There are about 30 Galactic \gr\ sources detected by EGRET.  Except for
the few identified as pulsars, the nature of most of them remains
uncertain.  The discovery of Geminga as a radio-quiet pulsar opened up the
possibility that additional pulsars might be detected in \gr s but not in
radio.  If \gr s are produced in the outer-magnetospheric accelerator,
they will have a large latitudinal coverage.  Because radio emission is
generally believed to be produced on the open field line bundle near the
polar cap, it has a narrow beam.  In this picture, there is a large
probability that the radio beam does not intersect the Earth while the
\gr\ beam does. Thus, the population of radio-quiet \gr\ pulsars is
potentially high enough to account for the majority of the unidentified
Galactic EGRET sources (\cite {hal93}; \cite{yad95}).  It is difficult to
discover periodicity in the \gr\ data because of the paucity of photons,
although this may be possible with future \gr\ missions such as {\it
GLAST}. The identification of Geminga was made from its X-rays, of which
the count rate is relatively high.  X-ray counterparts of high-energy \gr\
sources may be more generally useful in finding Geminga-like pulsars.  The
relative abundance of X-ray photons allows spectral analysis, and our
model predicts that a power-law spectrum in hard X-rays is associated with
a \gr\ pulsar.  This signature might provide important constraints in
selecting the prospective pulsar candidates. 

\acknowledgements
 
We thank K.~Chen, J.~Arons and E. Massaro for enlightening discussions. 
FYHW acknowledges support from NASA grant NAG 5-2524, JPH from NASA grant
NAG 5-3229, MR and TZ from NASA grant NAG 5-2841.  We also thank our
referee for his comments.  This paper is contribution 643 of the Columbia
Astrophysics Laboratory.

\clearpage

\begin{figure}
\plotone{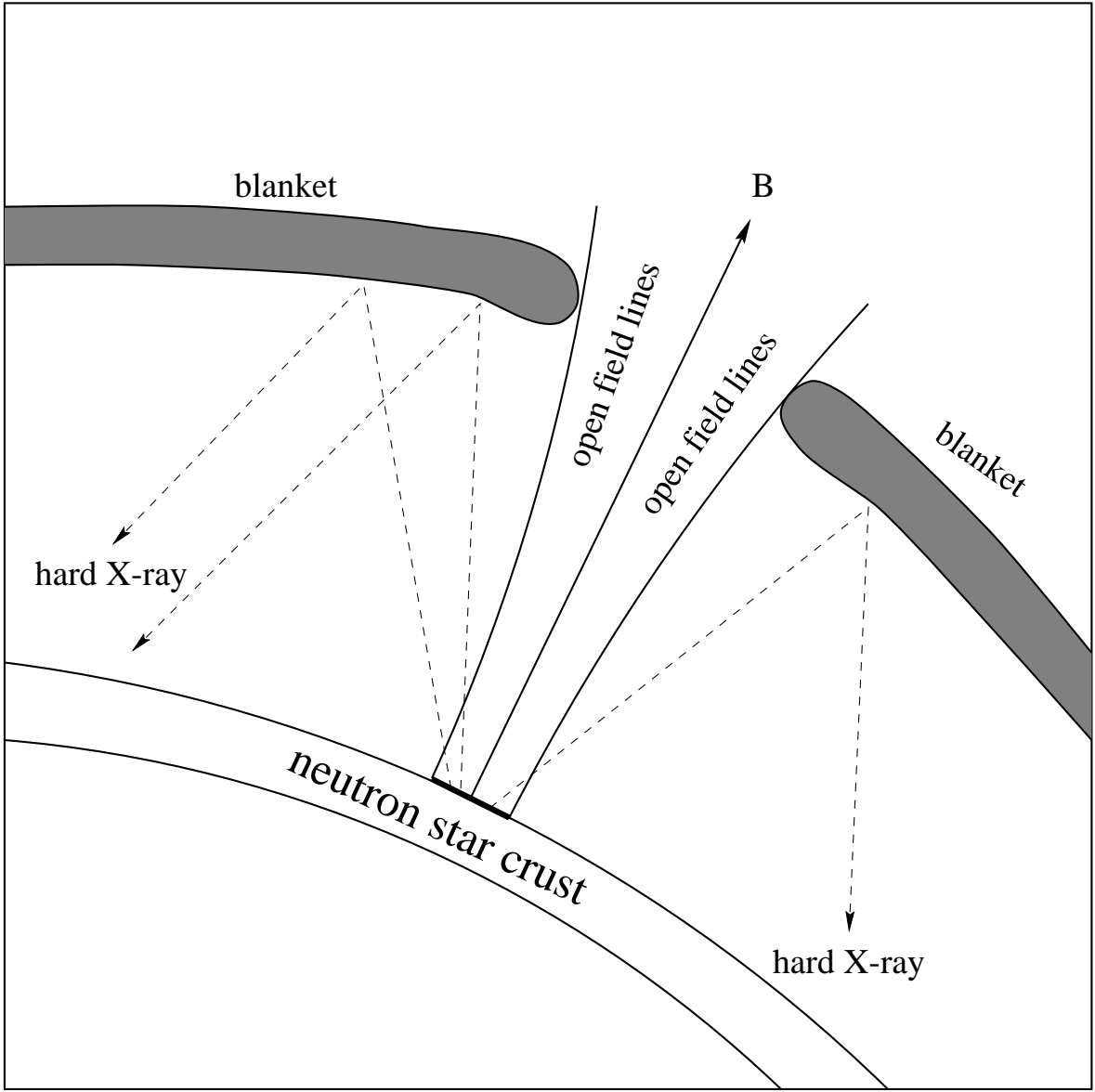}
\caption{For \gr\ pulsars, the \ep\ pairs created on closed field lines
will form a reflecting blanket around the neutron star; hard X-rays from
the heated polar caps will be reflected back to the surface of the star,
and a direct observation of them is prevented. \label{fig6}}
\end{figure} 

\begin{figure}
\plottwo{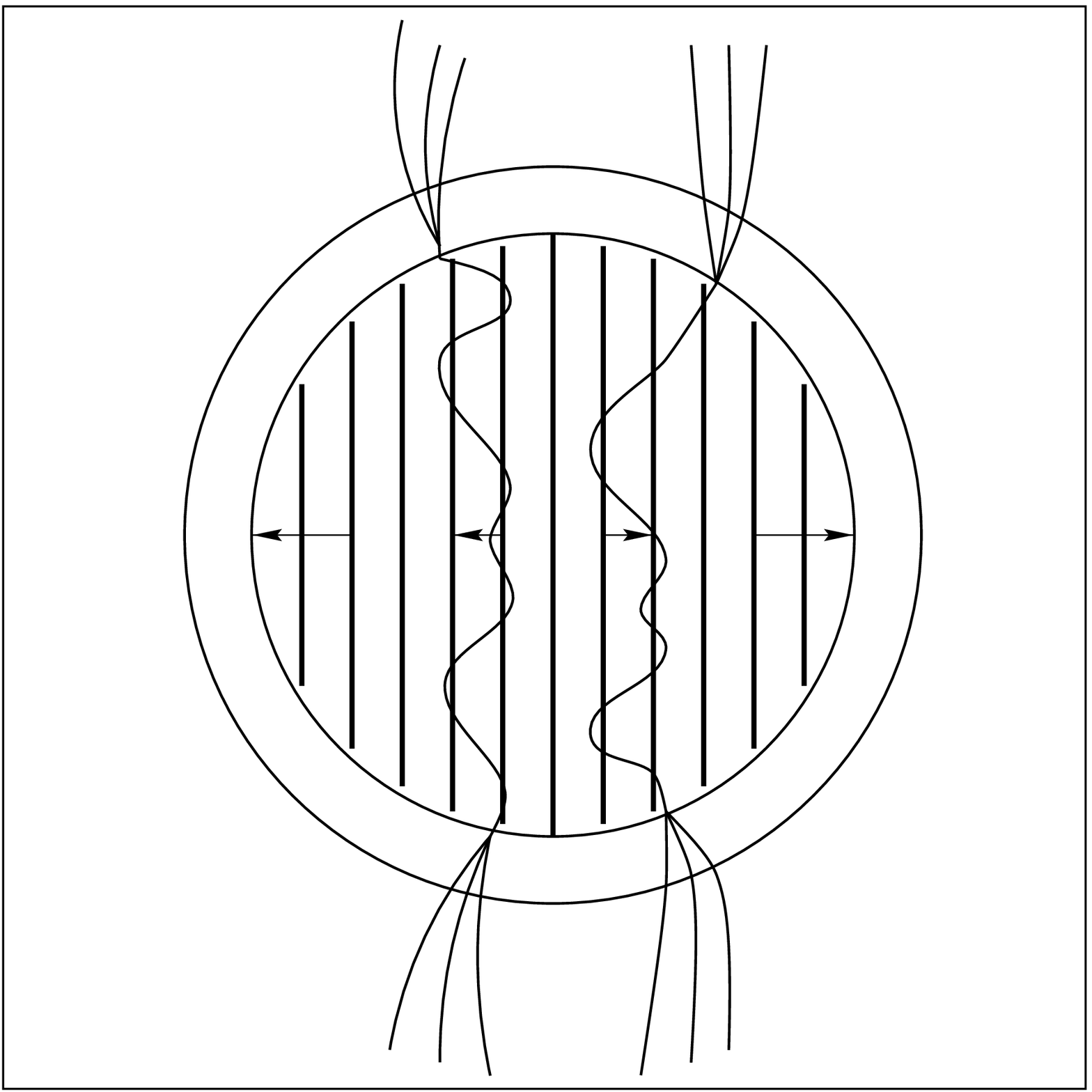}{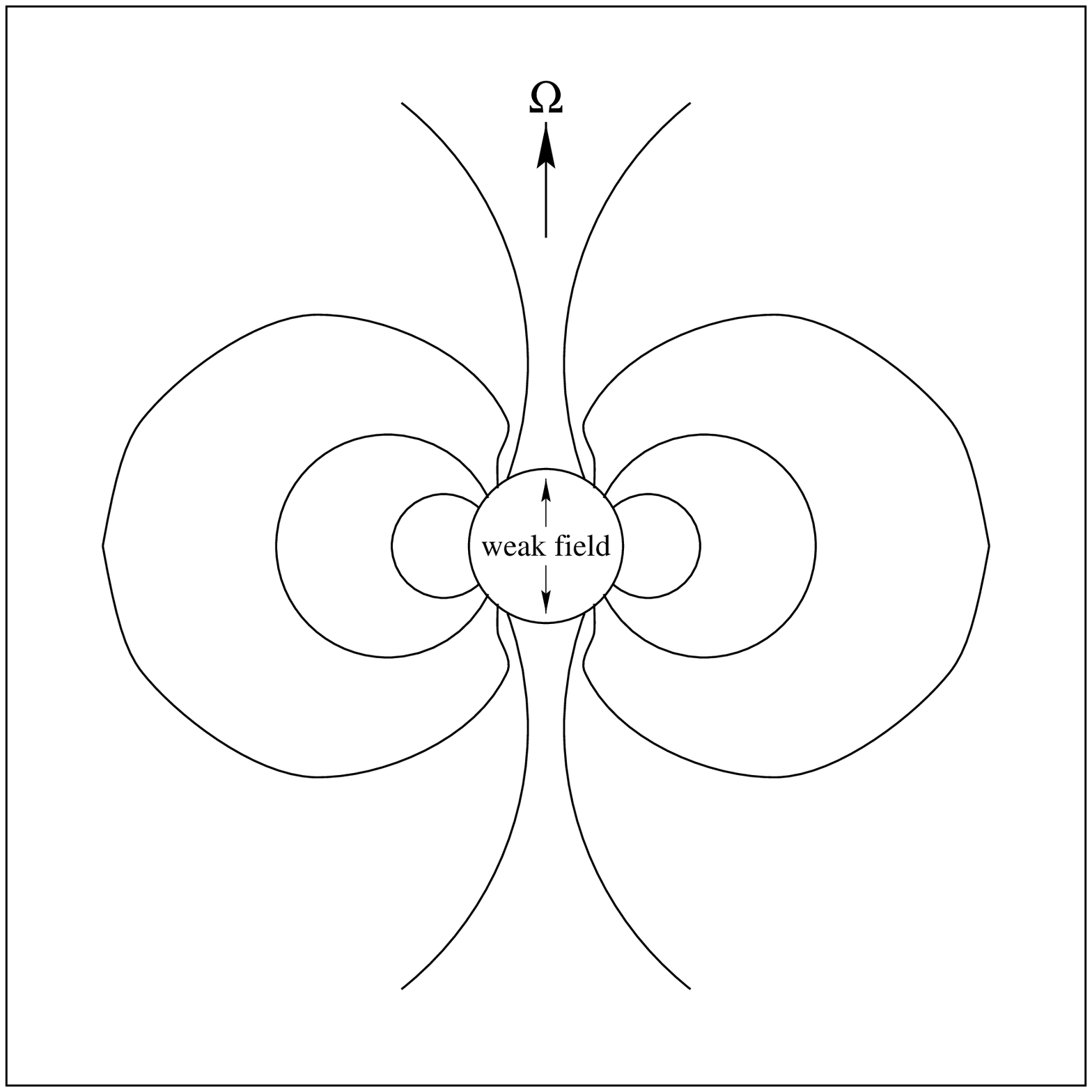}
\caption{(a) 
The interior of a neutron star consists of superfluid neutrons and
superconducting protons.  The quantized superfluid vortex lines are
parallel to the spin axis, but the superconducting flux tubes have a
complicated twisted toroidal and poloidal structure.  The
superfluid vortex line array expands as the neutron star
spins down. (b) The outward motion of vortex lines near the spin axis is
slow enough to carry core flux tubes and crust magnetic field with these
vortices and to leave a local region with anomalously weak magnetic field. 
The distorted field lines near the surface of the neutron star give a
polar cap area larger than would be the case for a central dipole.
\label{fig7}}
\end{figure}

\clearpage
 
\begin{deluxetable}{lrrrrr}
\footnotesize
\tablecaption{Pulsar Data \label{tbl-1}}
\tablewidth{0pt}
\tablehead{
\colhead{} & \colhead{Geminga} 
& \colhead{PSR 1055-52} &
\colhead{PSR 0656+14} & \colhead{PSR 1929+10}
& \colhead{PSR 0950+08}}
\startdata
$f$ (Hz) & 4.218 & 5.073 & 2.598 & 4.414 & 3.951 \nl
$\dot{f}$ (Hz s$^{-1}$) & $-1.95 \times 10^{-13}$  & $-1.5\times 10^{-13}$
&$-3.7\times 10^{-13}$ &$-2.25\times 10^{-14}$ &$-3.6\times 10^{-15}$ \nl
$I \Omega \dot{\Omega}$ (\lum) & $3.3 \times 10^{34}$ & $3.0 \times 10^{34}$ 
&$3.8\times 10^{34}$ &$3.9 \times 10^{33}$
&$5.6 \times 10^{32}$\nl
$\tau$ (year) &    3.4$\times 10^{5}$      & 5$\times 10^{5}$ 
&1.1$\times 10^{5}$ & 3.1 $\times 10^{6}$ & 1.7$\times 10^{7}$ \nl
$B_{\rm p}$ (10$^{12}$G)        & 1.6 & 1.1 & 4.7 & 0.5 & 0.24 \nl
$T_{1}$ (K) &   5.77$\times 10^{5}$ & 7.9$\times 10^{5}$ 
&8.1$\times 10^{5}$ & - & -  \nl
$L_{1}$\tablenotemark{a} \ (\lum) &1.47$\times 10^{31}$ 
& 2.3$\times 10^{32}$ &4.3$\times 10^{32}$ & - & - \nl
$\alpha$\tablenotemark{b} & $0.47^{+0.25}_{-0.23}$    
& $0.5^{+0.3}_{-0.3}$ & - & - & - \nl
$T_{2}$ (K) & - & - & 1.68$\times 10^{6}$ & 
$5.14 \times 10^{6}$  & $5.70 \times 10^{6}$  \nl
$L_{2}$\tablenotemark{c} \ (\lum)& 8.1$\times 10^{29}$  
& 1.5$\times 10^{30}$ &2.6$\times 10^{31}$ & 
$1.3 \times 10^{30}$ & $3.5 \times 10^{29}$  \nl
$A_{2}$\tablenotemark{d} \ ($\rm cm^{2}$) & - & - & $5.7 \times 10^{10}$
& $3.2 \times 10^{7}$ & $5.8 \times 10^{6}$ \nl
$d$ (pc) & 160 & 500 & 400 & 250 & 125 \nl 
\enddata
\tablenotetext{a}{Bolometric luminosity of the soft X-ray component}
\tablenotetext{b}{Power law energy index}
\tablenotetext{c}{Luminosity of the hard X-ray component; 0.7-5.0 keV for 
Geminga and PSR 1055-52, bolometric luminosity for PSR 0656+14,
PSR 1929+10, and PSR 0950+08}
\tablenotetext{d}{Fitted area of polar cap}
\tablerefs{\cite{hal97}; \cite{gre96}; \cite{wan97}}
\end{deluxetable} 
 
\end{document}